\documentclass[preprint,floats,eqsecnum,aps,tightenlines,showpacs]{revtex4}
\usepackage{epsfig}
\usepackage{color}

\begin{document}

\newcommand{\be}{\begin{equation}}
\newcommand{\ee}{\end{equation}}
\newcommand{\bea}{\begin{eqnarray}}
\newcommand{\eea}{\end{eqnarray}}
\newcommand{\nn}{\nonumber}
\newcommand{\dk}{\frac{d^4k}{(2\pi)^4}}
\newcommand{\ep}{i\epsilon}
\newcommand{\dq}{\frac{d^4q}{(2\pi)^4}}
\newcommand{\al}{\alpha}
\newcommand{\om}{\omega}

%

\title{Higgsonium in singlet extension of Standard Model \\
} 

\author{V.~\v{S}auli}
\affiliation{CFTP and Departamento de F\'{\i}sica,
Instituto Superior T\'ecnico, Av.\ Rovisco Pais, 1049-001 Lisbon,
Portugal, }
\affiliation{Department of Theoretical Physics,
Nuclear Physics Institute, \v{R}e\v{z} near Prague, CZ-25068,
Czech Republic}
\begin{abstract}
A possible formation of bound state is predicted in the Standard Model extension with
with additional scalar singlet. 
A suitable method of solution of the Bethe-Salpeter equation for a Higgsonium
bound state is proposed. With the help of an integral
representation the results are  directly obtained in Minkowski space. 
The appearance of Higgsonia is shown to be quite natural for considered extension of Standard model  
and the calculations are presented for the  bound state composed from  
two scalars which result from the large mixture of electroweak doublet and singlet eigenstate.
\end{abstract}

\maketitle

\section{Introduction}
The Standard Model (SM) of particle physics involves very minimalistic idea of electroweak symmetry breaking scenario. Consequently SM results with the single real scalar field representing the last   experimentally unobserved particle of SM - the Higgs boson. However, being inspired by the family pattern of SM fermionic sector, it is quite natural to consider an extensions of SM with more rich structure of the scalar sector. By doubling the doublets or/and adding the gauge singlets, such next-minimal extensions of SM model have been considered and studied from various perspectives. Clear motivation for such extensions is to reduce some SM shortcomings, like the better agreement with precision electroweak fit, the theoretical problem of mass hierarchy or the dark matter problem.
Adding the singlet scalar is worthstanding  in recently constrained SUSY models as well as \cite{DEGU07}.

Having more interacting scalar bosons, one can expect  qualitative changes in the scalar boson sector. In some circumstances the binding forces between scalars can appear strong enough to produce  bound states. What is the spectrum of appearing bound states and how they exhibit their existence in the collider experiments are important questions. 
To address this issue is in general a nonperturbative task and one has to make a reliable approximation. The later should be principally improvable to check the consistency of the assumptions made. To find a spectrum we assume that bound states are sufficiently stable to be decoupled from the continuum so they can be identified by the solution of homogeneous Bethe-Salpeter equation (BSE). 

To avoid extraordinary heavy states within strong couplings  a specific Higgs selfinteraction and extension of Higgs sector is required \cite{GRITRO07}. As we know the light Higgsonia could appear when 
more interacting scalars are present in the nature. So observing a Higgsonium resonance at LHC could be another evidence of the physics beyond the SM, whose details remains to be studied. 
As a simplest model we choose for our actual calculation is the extension of SM that involves the addition  of a real scalar singlet $S$ to the SM Lagrangian. The phenomenological implications for singlet extension  SM 
(xSM) has been studied from the collider and cosmological perspectives \cite{LHCscalar}. The later typically require small mixing with the SM Higgs and from the perspective of bound states
it reduces to the SM. In such circumstances it was shown in \cite{RUPP} that 
super heavy Higgs $m_H\simeq 1 TeV$ would be needed to form a bound state. In our model we will consider large mixing, which gives us two scalars $H_1$ and $H_2$ both having masses at few hundred $GeV$. Such model is a slight generalization (without $Z_2$ symmetry) of universal doublet-singlet Higgs couplings model \cite{BBN}.  Assuming that trilinear couplings have a little effect on electroweak precision observables $S$ and $T$ then the lighter scalar is experimentally restricted by the SM Higgs alone, i.e. its allowed mass is between $114 GeV$ and $145GeV$, while preferred value of heavier $H_2$ should lie in the range $\simeq 200- 250 GeV$ \cite{BBN}. In such circumstances, the most promising candidate for Higgsonium consist of two $H_2$ with a mass being not far form the sum of their on-shell masses. The main motivation of this paper is to find the explicit solution and figure out the conditions under which the Higgsonium is formed.

\section{xSM with large mixing of scalars}

The Lagrange density for the xSM  model is
\begin{equation}
{\cal L}=(D_{\mu}H)^{\dagger}D^{\mu}H+{1 \over 2}\partial_{\mu}S\partial^{\mu}S -V(H,S),
\end{equation}
where $H$ denotes the complex Higgs doublet and $S$ the real scalar. The term linear in $S$ is chosen to vanish after the spontaneous breaking. The potential is given by
\begin{eqnarray}
&&V(H,S)= {\lambda } (H^{\dagger}H-{v^2 \over 2})^2 +{\delta_1 \over 2}H^{\dagger}H~S  \\
&&+{\delta_2 \over 2}H^{\dagger}H~S^2+ {\delta_1 v^2} S+{\kappa_2 \over 2} S^2 +{\kappa_3 \over 3}S^3+{\kappa_4 \over 4}S^4. \nonumber 
\end{eqnarray}

In unitary gauge the charged component of the Higgs doublet $H$ becomes the longitudinal component of the charged $W$-bosons and the imaginary part of the neutral component becomes the longitudinal component of the $Z$-boson. In unitary gauge the Higgs field  doublet reads 
\be
\left({\begin{tabular}{c}
   $ 0 $   \\
$\frac{1}{\sqrt{2}}(h+v)$\\   
\end{tabular}}\right) \, .
\ee
Within our notation the mass terms in the scalar potential become
\be
V_{ \rm mass}={1 \over 2}\left(\mu_h^2 h^2 +\mu_S^2 S^2 +\mu_{h S}^2 h S \right),
\ee
where
\begin{eqnarray}
\label{param}
&&\mu_h^2=2\lambda v^2 \nonumber \\
&&\mu_S^2= \kappa_2+\delta_2 v^2/2  \nonumber \\
&&\mu_{h S}^2=\delta_1 v.
\end{eqnarray}
The mass eigenstate fields $H_{1,2}$  are linear combinations of the Higgs scalar field $h$ and the singlet scalar field $S$. Explicitly, the inverse transformation reads
\bea
h&=& c ~H_1- s ~H_2\, ;
\nn \\
S&=& s ~H_1+ c ~H_2\, ,
\eea
where $c=cos \theta$, $s=sin \theta$  and the mixing is determined as

\begin{equation}
\tan \theta={ x \over 1+\sqrt{1+x^2}},~~~x={\mu_{h S}^2 \over \mu_S^2 - \mu_h^2 }.
\end{equation}
for $x>1$. For a heavier singlet $(x<0) $ we have the mixing angle
\be
\tan \theta={   1+\sqrt{1+x^2} \over |x|} \, .
\ee
Note, here we differ from the convention of ~\cite{KRASNIKOV,KORAWI2006}
by the sign of $x$. 

The small $\delta_1$ mixing scenario with the light singlet like scalar has been considered in 
\cite{KORAWI2006}.  For such a case we do not  expect bound state since the triplet interaction between scalars is very weak.
The terms in the scalar potential that break the discrete $S \rightarrow
-S$ symmetry are proportional to the couplings $\delta_1$ and $\kappa_3$ and  
we do not consider these terms as very small ones, instead we assume they are large enough to make a bound state providing also sufficiently strong communication with the rest of the SM.

In this paper we will consider relatively large mixing angle $\theta$ with  Higgs masses close to few hundred GeV which could be  promising for experimental observation thorough the LHC era. For such case the constraints from electroweak precision observables and their implications for the LHC Higgs phenomenology has been already analyzed in \cite{LHCscalar} .

The mass eigenstates are in any case
\bea
M_{1}^2&=&\mu_h^2 c^2 +\mu_S^2 s^2+ \mu_{hS}^2 c s  
\nn\\
M_{2}^2&=&\mu_h^2 s^2 +\mu_S^2 c^2- \mu_{hS}^2 c s \, .
\eea

The rest of the scalar potential  contains four new parameters which are added to the SM.
\be
V_{int}=\frac{\lambda}{4}h^4+\frac{\kappa_4}{4}S^4+\lambda v h^3+ \frac{\kappa_3}{3}S^3
+\frac{\delta_2}{2}h^2S^2+\frac{\delta_1}{2}h^2S+\frac{\delta_2}vh S^2
\ee
Based on the nonrelativistic consideration,  the trilinear interaction should be sufficiently enhanced against the quartic one, otherwise the bound state cannot be formed. 

The purely cubic interaction between mass eigenstates can be written in the following way
\be
V_{cub}=g_{111}H_1^3+g_{112}H_1^2H_2+g_{122}H_1H_2^2+g_{222}H_2^3 \,
\ee
with the new couplings    related to the old ones as
\be
\left({\begin{tabular}{c}
   $ g_{111} $   \\
$ g_{112}$\\   
$ g_{122}$\\
$ g_{222}$\\
\end{tabular}}\right)=
\left({\begin{tabular}{cccc}
   $ s^3 $ & $ c^3$&$sc^2$& $s^2c$   \\
$ 3s^2c$ & $ -3sc^2$ & $(c^3-2s^2c)$ & $(2sc^2-s^3)$\\
$ 3sc^2$ & $ 3s^2c$ & $(s^3-2sc^2)$ & $(c^3-2s^2c)$\\
$ c^3$ & $ s^3$ & $s^2c$ & $-sc^2)$\\
\end{tabular}}\right)
\left({\begin{tabular}{c}
   $ \frac{\kappa_3}{3} $   \\
$ \lambda v$\\   
$ \frac{\delta_1}{2}$\\
$ \delta_2 v$\\
\end{tabular}}\right)
\ee

The quartic couplings between the physical states are related to the original Lagrangian 
parameters in the following way
\be
V_{4}=\frac{g_{111}}{4}H_1^4+\frac{g_{1112}}{4}H_1^3H_2+\frac{g_{1122}}{4}H_1^2H_2^2+\frac{g_{1222}}{4}H_1H_2^3+\frac{g_{2222}}{4}H_2^4
\ee
\be
\left({\begin{tabular}{c}
   $ \lambda_{1111} $   \\
$ \lambda_{1112}$\\   
$ \lambda_{1122}$\\
$ \lambda_{1222}$\\
$ \lambda_{2222}$\\
\end{tabular}}\right)=
\left({\begin{tabular}{ccc}
$ c^4$ & $ s^4$ & $s^2c^2$\\
$ -4sc^3$ & $ 4s^3c$ & $2sc(c^2-s^2)$\\
$ 2s^2c^2$ & $ 2s^2c^2$ & $(1-6s^2c^2)$\\
$ -4s^3c$ & $ 4sc^3$ & $2sc(s^2-c^2)$\\
$ s^4$ & $ c^4$ & $s^2c^2$\\
\end{tabular}}\right)
\left({\begin{tabular}{c}
   $ \lambda $   \\
$ \kappa_{4}$\\   
$ 2\delta_2$\\
\end{tabular}}\right) \, .
\ee

\section{Higgsonium in Quantum Field Theory}

Higgsonium, being a two body bound states of two scalars is described by two body BSE:
\be
\Gamma=\int_k VG^{[2]}\Gamma \,
\ee
where we use shorthand notation $\int_k=i\int\frac{d^4q}{(2\pi)^4}$ and 
$G^{[2]}$ is the two particle propagator of the constituent Higgsies, 
just for now let us consider Higgsonium made of two $H_1$s.
In momentum space it can be conventionally written as
\bea
G^{[2]}(k,P)&=&D(k+P/2,M_1^2)D(-k+P/2,M_1^2) \,  ;
 \\
D(k,M^2)&=&\frac{1}{k^2-M^2-\ep} \, .
\eea

Let assume that the attractive interaction between heavy Higgsies $H_1$ is
strong enough to form a bound state. Within the xSM  the irreducible BSE kernel in lowest order reads
\be \label{tree}
V=6\lambda_{1111}+\sum_{x=s,t,u}\left[\frac{4 g_{112}^2}{x-M_2^2}+\frac{36 g_{111}^2}{x-M_1^2}\right] \,
\ee
where the first term represents a pure constant interaction and the $s,t,u$ are the usual Mandelstam variables.

In the next it is advantageous to explicitly divide the solution which is independent on the relative momentum
of the constituents. Following the notation in \cite{RUPP} the original BSE can be rewritten in the following form
\be \label{BSE}
\Gamma_p(p,P)=\Gamma_I(P)\int_k V_p(k,p,P) G^{[2]}(k,P)+\int_k V_p(k,p,P) G^{[2]}(k,P)\Gamma_p(k,P)\, ,
\ee
where  
\bea
 V_I=V_c+V_s \, , \, V_p&=&V_t+V_u \, ,
\eea
(to avoid confusion with the particle content, here we use the letter $I$ instead of index $1$ and $p$ instead of $2$ originally used in the paper \cite{RUPP}) The first term  collects  all the constant term, i.e the ones that do not depend on the relative momentum.  In our tree level kernel approximation it  reads
\be \label{const}
V_I=6\lambda_{2222}+\frac{4 g_{221}^2}{P^2-M_1^2}+\frac{36 g_{222}^2}{P^2-M_2^2} \, .
\ee
So the full solution of BSE is given by the sum
\be
\Gamma(p,P)=\Gamma_I(P)+\Gamma_p(p,P) \, ,
\ee
where the equation for  the function $\Gamma_I(P)$ is purely algebraic
\be  \label{druha}
\Gamma_I(P)=\frac{V_I \int_k  \Gamma_p(k,P)G^{[2]}(k,P)}{1-V_I\int_k G^{[2]}(k,P)}
\ee

The BSE represents the singular equation which can be solved by some known method.
One known possibility is to perform a Wick rotation for a relative momenta of constituents while keeping the total square of four momenta $P^2$ timelike.

The other well known possibility is the Minkowski solution performed within the utilization  of the known integral representation for the kernels and amplitudes that appear in the BSE.
Then the momentum space BSE can be rewritten to the equivalent integral equation for the weight function. This, historically named Perturbation Theory Integral Representation \cite{NAKAN}
has been successfully utilized to solve relativistic bound state problems in various models and approximations \cite{NAK2,S1,S2,S3,KUSIWI97}. We straightforwardly apply the method here. 

The bound state vertex function can be expressed as    
\bea \label{bounstate}
\Gamma(P,p)=\int\limits_{-1}^1 d\eta\,
\int\limits_{-\infty}^{\infty}d\alpha\,
\frac{\rho^{[n]}(\alpha,\eta)}{[F(\al,\eta;P,p)]^n} \, ,
\eea
where $n$ is an arbitrary integer and all the singularities are trapped by the zeros of the denominator in (\ref{bounstate}) which reads 
\be
F(\al,\eta;P,p)=\al-(p^2+P.p z+\frac{P^2}{4})-\ep
\ee

 Here, having the quartic interaction present, the inhomogeneous term is generated represented by $\Gamma_I(P)$ which is just real constant for a given discrete value of bound state mass $P$.
So far all  delta distributions in full $\rho$ should be factorized and integrated out, thus we naturally take
\bea \label{bound2}
\Gamma(P,p)=\Gamma_I(P)+\int\limits_{-1}^1 d\eta\,
\int\limits_{-\infty}^{\infty}d\alpha\,
\frac{\rho_p(\alpha,\eta)}{[F(\al,\eta;P,p)]} \, ,
\eea
where $\rho_p(\alpha,\eta)$ is assumed to be a real function, and not a delta distribution. The nontrivial part of integral representation corresponds exactly with the function $\Gamma_p$, noting its structure is fully driven by pure triplet interaction of the Higgs. Furthermore, we explicitly choose $n=1$ in the integral representation  (\ref{bound2}), following 
the most easy integral representation of the inhomogeneous term 
in the expression, i.e.  $\int_k V_p(k,p,P) G^{[2]}(k,P)$. In fact, These integrals correspond to the  scalar triangle Feynman diagram. Indeed,  written more explicitly, we get
\bea
\int_k V_p(k,p,P) G^{[2]}(k,P)&=&
4g_{112}^2\int_k \left[ D(k-p,M_2^2)+D(k+p,M_2^2)\right] G^{[2]}(k,P)
\nn \\
&+&36g_{111}^2\int_k \left[ D(k-p,M_1^2)+D(k+p,M_1^2)\right] G^{[2]}(k,P)
\nn \\
&=&8 g_{112}^2 I_{\Delta 112}(P,p)+72 g_{111}^2 I_{\Delta 111}(P,p)
\nn \\
&=&\int\limits_{-1}^1 d z\,
\int\limits_{-\infty}^{\infty}d\alpha\,
\frac{8 g_{112}^2\rho_{\Delta 112}(\alpha,z)+72 g_{111}^2\rho_{\Delta 111}(\alpha,z)}{(\alpha-M_1^2)F(\al,z;P,p)}\, ,
\eea
where the functions $\rho_{\Delta}$s are listed in the Appendix $A$ in Rel. (\ref{rhoI}).
The factor $(\alpha-M_1^2)$ in the last line typically factorizes  and here it is  the matter of our convenient convention.
 
From human effort  point of view the choice $n=1$ represents most convenient one, however recall here the known  property of superrenormalized model studied yet: The function $ \rho^{[n]}$   is more smooth for larger $n$. The BSE were solved for generalized Wick-Cutkosky models for the lowest value  $n=1,2$ in practice (the exception is the original Wick model, where $n$ labels different energy states and is not arbitrary at all, this is because of  massless boson). The later choice usually lead to meaningful improvement of numerical accuracy. In our case with $n=1$ we expect 1-2 percentage accuracy in $P^2$ eigenvalue identification.

In order to express the second term in (\ref{BSE}) we  use the integral representation for  $\Gamma_p$, hence we can write
\bea
\int_k V_p(k,p,P) G^{[2]}(k,P)\Gamma_p(k,P)&=&\int\limits_{-1}^1 d z\,
\int\limits_{-\infty}^{\infty}d a \, 
\rho_2(a,z)\times
 \\
&&\left\{4g_{112}^2\left[I_{t112}(P,p;a,z)+I_{u112}(P,p;a,z)\right]\right.
\nn \\
&&+\left.36g_{111}^2\left[I_{t111}(P,p;a,z)+I_{u111}(P,p;a,z)\right]\right\}\, ,
\nn
\eea 
where

\bea
I_{t112}(P,p;a,z)&=&\int_k \, D(q+P/2,M_1^2)D(-q+P/2,M_1^2)\frac{D(p-q,M_2^2)}{F(a,z;P,q)}\, , 
\\
I_{u112}(P,p;a,z)&=&\int_k \, D(q+P/2,M_1^2)D(-q+P/2,M_1^2)\frac{D(p+q,M_2^2)}{F(a,z;P,q)}\, ,\nn
\eea
and the  function $I_{t111}$, $I_{u111}$ can be obtained from $I_{t112}$, $I_{u112}$ through the simple change $M_2\rightarrow M_1$ . 

The functions $I_{t},I_{u}$ satisfy the following integral representation  
\bea
I_{t,u}(P,p;a,z)&=&  \int\limits_{-1}^1 d\eta\,
\int\limits_{-\infty}^{\infty}d\alpha\,
\frac{\rho_{t,u}(\alpha,\eta,a,z)}{(\alpha-M_1^2)F(\al,\eta;P,p)} \,
\eea
where we have suppressed the propagators indices for brevity and
where the functions $\rho_{t,u}$ are reviewed in the Appendix \ref{leading}.
As simple consequence of crossing symmetry we have 
\be
\rho_{u}(\alpha,\eta,a,z)=\rho_{t}(\alpha,-\eta,a,z) \, .
\ee

At this point we can assume the  validity of  theorem of uniqueness derived for general PTIR
which implies that the original BSE is equivalent to the solution of the following regular real integral  equation:
\be \label{vysledek}
\rho_p(\alpha,\eta)= \frac{1}{\al-M_1^2}\left[\Gamma_I(P)\rho_{I}(\alpha,\eta)+
 \int\limits_{-1}^1 d z\, \int\limits_{-\infty}^{\infty}d a \,\rho_p(a,z) 
{\cal V}(\alpha,\eta,a,z)\right] \, ,
\ee
where 
\bea 
\rho_{I}(\alpha,\eta)&=&8g_{112}^2\rho_{\Delta 112}(\alpha,\eta)+
72g_{111}^2\rho_{\Delta 111}(\alpha,\eta) \, ,
\nn \\
{\cal V}(\alpha,\eta,a,z)&=&4g_{112}^2\left[\rho_{t112}(\alpha,\eta,a,z)+\rho_{t112}(\alpha,-\eta,a,z)\right]
\nn \\
&+&36g_{111}^2\left[\rho_{t111}(\alpha,\eta,a,z)+\rho_{t111}(\alpha,-\eta,a,z)\right]\, .
\eea
As a consequence of the quartic constant interaction term  Eq. (\ref{vysledek}) is inhomogeneous
and necessarily coupled  to the Eq. (\ref{druha}). Using the integral representation for $\Gamma_p$ 
\be
\Gamma_I(P)=\frac{V_I^R  \int\limits_{-1}^1 d z\, \int\limits_{-\infty}^{\infty}d a
\rho_p(a,z) I_{F}(P^2;a,z)}{1-V_I^R I_B^{[R]}(P^2)} \, ,
\ee
where $V_I^R$ is the renormalized constant interaction and the  suitable expression for the loop integrals $I_F$ and $I_B$ are reviewed in the Appendix \ref{buble}. 

In this work we use the momentum subtraction renormalization scheme with zero momentum scale.
Hence
\be
V_I^R=6\lambda_{1111}^R+\frac{4g_{112}^2}{P^2-M_2^2}+\frac{36g_{111}^2}{P^2-M_1^2}
\ee
where $\lambda_{1111}^R$ is the renormalized quartic coupling of heavier Higgs mass eigenstates
$H_1$, the others couplings stem from superrenormalizable interaction, hence they do not need to be renormalized at all. For a fixed bound state mass $P^2$ the function  $\Gamma_I(P)$ is a simple constant, affecting  effectively the strength of the inhomogeneous term. Note that there is a peak for $I_B^{[R]}(P^2)$ when approaching the continuum and the one loop effect can be quite large.

\section{Results}

The inhomogeneous BSE requires renormalization since the function  $\rho_p$ is linearly present at each term of BSE rhs. After a suitable normalization then the BSE for the weight function (\ref{vysledek}) has been solved by the method of iterations. The coupling constants have been varied to reach the real discrete spectrum of Higgsonia.

Firstly, let us mention the result for the SM Higgsonium as it follows from our method of BSE and compare with \cite{RUPP} . In this case, the only experimentally constrained input is the Higgs vev. $v=275/\sqrt{2} GeV$ while the Higgs mass and the cubic coupling  depends on presently unknown 
coupling $\lambda$, these satisfy $m_h=\sqrt{2\lambda}v, \lambda_3=2\lambda v$. As it is known, there are no bound states below a certain critical coupling of $\lambda_3$. The first state appears with the mass close to the on shell masses of Higgses $M=2m_h$ ($M=1.95m_h$ was taken in practice) and it is formed when  $m_h=1.28 TeV$. This value is in a reasonable agreement with \cite{RUPP}, however the input mass of such Higgs is  ruled out by electroweak precision measurements. 
 Furthermore, for such a fat Higgs, the Higgs sector of SM represents strongly coupled field theory  and our BSE solution becomes only a rough estimate. In addition, being more realistic and  switching on the top quark Yukawa, such fat Higgs becomes broad resonance and its fast decays should  prevent the formation for bound states. The inhomogeneous BSE alone turns out to be not well controlled tool at all in this case.

To achieve the numerical solution of Higgsonium in  xSM we have firstly adjusted the couplings between the gauge eigenstate to get  physical masses and couplings between the the mass eigenstates
$H_1,H_2$ and then we look for the BSE solution varying the mass $P^2$. A more convenient approach where the bound state mass has been fixed and some of the couplings have been varied has been followed as well. Both approaches have been found equivalent, however  feasibility of getting the numerical solution can be different.  The later approach, if needed,  requires re-identification of the original gauge eigenstates couplings. 

As we can not scan the all parameter space,  we choose some exemplatory group, we suppose  is relevant for calculation of a bound states. An example of  the parameters in used as an input is as the following: $v=275 GeV, \, \lambda=0.20, \, \delta_1=1.20*v, \, \delta_2=0.40$,
$\kappa_2=0.10*v, \,   \kappa_3=5.0*v , \, \kappa_4=0.20$. It leads to two massive eigenstates with $M_1=179.5 GeV, \, M_2=177.7 GeV,$ and couplings $g_{111}\simeq 280 GeV,g_{222}\simeq 400 GeV $ , 
$\lambda_{1111}\simeq 0.33$ and the appropriate mixing is 
$cos\theta=0.696$, such large value of mixing is physically justified since the masses $M_1,M_2$ 
are in the range allowed by electroweak precision test \cite{LHCscalar}. In this case we found the solution of BSE that produces 
the bound states 20 $\%$ lighter than the Higgs production threshold :
\be
M_B=1.6\times M_1=286 GeV\, . \nn
\ee

In general it is clear that the trilinear couplings must be large enough to ensure bound states, otherwise we get only continuum spectrum of free Higgses in a production experiments.
As expected and observed, increasing trilinear couplings  the mass of the bound state decrease and  goes away from the  sum of on-shell constituent masses.  

At given stage, being limited by numerical accuracy and convergence,  we did not perform ultimate search for the limiting case $P=2M_{H1}$  solution, neither we were looking for today physically not well motivated bootstrap -deeply bounded state solutions $M_B<<(M_1+M_2)$. Instead of, we simplify situation and we kept the masses of Higgses fixed and vary only trilinear couplings $g_{111},g_{222}$
by a common prefactor $C$, $g_{111},g_{222}\rightarrow C g_{111},C g_{222}$ in the parameter space of mass eigenvalue physical states.
In this way we have found trilinear couplings have changed by the significant prefactors $C=0.41$ ($g_{111}=114.8 GeV,g_{222}=164 GeV $) in order to get the following bound state mass eigenvalue:
\be
 M_B=0.95*2 M_1=95.53\% (M_1+M_2) \, , \nn
\ee
 which value is already rather closed to the threshold. In other words: we predict weakly bounded Higgsonia in the singlet extension of SM with natural trilinear couplings (measured in units of Higgs masses), while the deep bound states are possible only when the coupling are un-naturaly tuned to the artificial and perhaps un-natural large values. Increasing the mass $M_B$ towards the threshold  we still expect some substantial decrement of trilinear couplings, however the $BSE$ did not provided stable and arbitrarily accurate solutions to the date.

In general, the lighter Higgs we have and for the  bound state mass  closed  to the thresholds $M_1+M_2$
we can expect that the presented form of the BSE is built up already in quite reliable approximations. Being not so far from the threshold, the next order  irreducible diagrams (e.g. cross diagrams etc.) contribution is getting  small since we stay effectively in the weak coupling regime (note that higher orders are suppressed by powers of $1/(4\pi)^2$). We do not present the BSE solutions for very deep bound state, which in principle can be interesting issue of pure numerical interest.  For a very deeply binded Higgs  we have no justification of approximations we made.
Also, as we can learn from QCD-like studies, the importance of dressing constituent (via Schwinger-Dyson studies) could be addressed for the strong  coupling systems. This issue is happily much less urgent for a weakly bounded Higgsonia, which as our study  shows up,  much likely exist in the singlet extension of SM.

\section{Conclusion}
Using a BSE formalism  we performed a simple search for scalar bound states in the  Higgs sector of SM and mainly in xSM extension of SM. In the second case, in order to have a reasonable model which is not completely  ruled out by electroweak oblique correction constraints we have assumed reasonably light scalars with large mixing. In this model, there are no light deeply  bounded states unless the new cubic coupling is  very large. On the other side, we predict bound state of the mass compared to the sum of the masses of the constituents, i.e. approximately non-relativistic bound states.    This is the  main  conclusion from the numerical inspection of assumingly most promising regime of xSM parameter space. The Higgsonia productions can be expected 
in xSM with large mixing and in similar models with more singlets and doublets as well. Our fully relativistic calculation shows that  Higgsonia can  appear in simple realistic  extensions of SM.  If some sort of Higgsonia will appear realized in the nature,   then the method presented here is improvable by considering improved kernels of the model. For instance,  taking into account the  selfenergy contributions  or considering more complicated cross  exchanges of Higgses is  straightforward through the perturbative introduction of appropriately known integral representations of such kernels  \cite{S2},\cite{CAKA2006}.

\begin{center}{ \bf Acknowledgments}\end{center}
I thank for discussion to G. Rupp who has brought my 
attention to this interesting problem.

--------------------------------------------------------------

\appendix

\section{Integral represenation for $I_{\Delta}(P,p)$}
\label{integrals}

In this appendix we review the integral representation for the following
scalar triangle diagram 
\be
 I_{\Delta}(P,p)=i\int\frac{d^4q}{(2\pi)^4}\,
D(q+P/2,M_1^2)D(-q+P/2,M_1^2)D(q-p,M_2^2) \, ,
\ee

This is just the scalar triangle diagram, hence the expression is finite and do not require any renormalization. The integral representation  is easily derivable by following the Feynman parametrization tricks and 
we refer the reader to the Appendix G of the work \cite{SAULIphd} for the details.

The amplitude reads $I_\Delta$  
\bea
I_{\Delta}(P,p,M_i)&=&  \int\limits_{-1}^1 d\eta\,
\int\limits_{-\infty}^{\infty}d\alpha\,
\frac{\rho_{\Delta}(\alpha,\eta)}{(\al-M_1^2)F(\al,\eta;P,p)} \, ,
\label{idelspec}\\
\rho_{\Delta}(\alpha,\eta)&=& -\frac{1}{32\pi^2}\, \sum_{i=\pm}
\frac{\Theta(t_i)\Theta(1-t_i)\Theta(D)}
{|\al-\frac{M_2^2}{t_i^2}- S|} \, ,
\label{rhoI}\\
t_{\pm}&=& \frac{\al+M_2^2-M_1^2\pm\sqrt{D}}{2(\al-S)} \, ,
\label{tpmsol}\\
\nn \\
 D&=&(\al-M_2^2-M_1^2)^2-4M_2^2(M_1^2-S) \, ,
\label{detD}
\eea
where we remind
\be
S=\frac{P^2}{4}(1-z^2)
\ee

Further note that, one can derive the following useful constraint
\be
 \alpha \geq \alpha_{min}=M_2^2+M_1^2+
  2 \sqrt{M_2^2 (M_1^2-S)} \, .
\ee
This can be utilized when enforcing the convergence of the numerical solution.

\section{ One loop bubble $I_B$ and  $I_{F}$}
\label{buble}

The one bubble integral $I_B$ reads
\be
 I_{B}(P^2)=i\int\frac{d^4q}{(2\pi)^4}\,
D(q+P/2,M_1^2)D(-q+P/2,M_1^2) \, ,
\ee
which is log divergent and so requires subtraction. In momentum subtraction scheme it defines the value of a renormalized quantity (coupling constant in our case) at the momentum scale $p^2=\mu^2$.
The renormalized one loop correction can be expressed as
\bea
I_{B}^{[R]}(P^2,\mu^2)&=&I_{B}(P^2)-I_{B}(\mu^2)
\nn \\ 
&=&\int_{4M_1^2}^{\infty}d\omega\frac{\sqrt{1-4M_1^2/\omega}}{P^2-\omega}
\frac{P^2-\mu^2}{\omega-\mu^2} \, ,
\eea
noting that the function $I_{B}$ is real for underthreshold momenta $P^2<4M_1^2$ which is the  relevant regime for our bound state study.

The contact term leads to the following type of the integral in BSE 
\be \label{sasko}
 I_{F}(P^2;a,z)=i\int\frac{d^4q}{(2\pi)^4}\,
D(q+P/2,M_1^2)D(-q+P/2,M_1^2)\frac{1}{F(a,z;P,q)}\, ,
\ee
This  bubble-like contribution  is finite since softened by the factor $1/F$ which stems from the bound state integral representation. The integral can be solved analytically by the  manner described bellow.

First let us use Feynman trick and match the propagator together, explicitly written we have
\be
  D(q+ P/2,M_1^2) D(-q+P/2,M_1^2)=
 \frac{1}{2} \int\limits_{-1}^1
 \frac{d \eta}{[F(M_1^2,\eta;q,P)]^2} \, ,
\ee 
matching further with the third fraction  in the expression  (\ref{sasko}) we get
\be
  \frac{D(q+ P/2,M_1^2) D(q-P/2,M_1^2)}{F(a,z;P,q)}=
\int\limits_{0}^1 dy \int\limits_{-1}^1 d \eta
\frac{-y}{[k.P(\eta-z)y+(\alpha-m^2)y+k^2+k.Pz+\frac{P^2}{4}-\alpha]^3} \, .
\ee 
Integrating over the momenta we obtain
\be
I_{F}(P^2;\alpha,z)=\frac{1}{2(4\pi)^2(\alpha-M_1)^2}\int\limits_{0}^1 dy \int\limits_{-1}^1 d \eta
\frac{-y}{\frac{P^2}{4}(1-\beta^2)-m^2y-\alpha(1-y)+\ep}\, ,
\ee
where $\beta=(\eta-z)y+z$.
After the substitution $\eta \rightarrow \beta $, the integration over the variable $y$ and some trivial algebra we can get the following expression 
\be
I_{F}(P^2;\alpha,z)=\frac{1}{2(4\pi)^2(\alpha-M_1^2)}
\left[\int\limits_{z}^1 d\beta\, ln\frac{u}{\frac{1-\beta}{1-z}+u}
+\int\limits_{-1}^z d\beta \, ln\frac{u}{\frac{1+\beta}{1+z}+u}\right]\, ,
\ee
where we defined
\be
u=\frac{M_1^2-\frac{P^2}{4}(1-\beta^2)}{\alpha-M_1^2}\, .
\ee

\section{Integral representation for $I_{t}$ and $I_{u}$}
\label{leading}

The leading term in the BSE  is a weighted sum over the following integral:
\be \label{svycarsko}
I_{t}(P,p;a,z)=i\int\frac{d^4q}{(2\pi)^4}\,
D(q+P/2,M_1^2)D(-q+P/2,M_1^2)\frac{D(p-q,M_2^2)}{F(a,z;P,q)}\, .
\ee
The appropriate expression is derived in the paper \cite{S2} and \cite{SAULIphd}.
Note, there are few misprints thorough the derivation in the original paper and for detailed derivation we recommend the later work \cite{SAULIphd}.
   

%
\bea
I_{t}(P,p;a,z)&=&  \int\limits_{-1}^1 d\eta\,
\int\limits_{-\infty}^{\infty}d\alpha\,
\frac{\rho_{t}(\alpha,\eta,a,z)}{(\al-M_1^2)F(\al,\eta;P,p)} \, ,
\\
\rho_{t}(\alpha,\eta;a,z)&=& \frac{1}{2(4\pi^2)}\,\sum_T  \sum_{i=\pm}
\frac{\Theta(x_i(T))\Theta(1-x_i(T))\Theta(D(T))}
{|\al-\frac{M_2^2}{x_i^2(T)}- S|} \, ,
\\
x_{\pm}(T)&=& \frac{\al+M_2^2-R(T)\pm\sqrt{D}}{2(\al-S)} \, ,
\\
\nn \\
 D(T)&=&(\al+M_2^2-R(T))^2-4M_2^2(\al-S) \, ,
\eea
where 
\be
R(T)= T a+(1-T) M_1^2\, ,
\ee
and the symbol $\sum_T$ is shorthand notation for the following sum
\be
\sum_T f(T)=f(0)-\theta(z-\eta)f(T_+)-\theta(\eta-z)f(T_-) \, ,
\ee
where $f$ is an arbitrary function and 
\be
T_{\pm}=\frac{1\pm\eta}{1\pm z}.
\ee

The term that is generated  due to the $u-term$  gives only a subleading contribution. Its derivation  can be very easily proceed by 
considering the change   $t\rightarrow u$. Explicitly we have 
\be \label{lipsko}
I_{u}(P,p;a,z)=i\int\frac{d^4q}{(2\pi)^4}\,
D(q+P/2,M_1^2)D(-q+P/2,M_1^2)\frac{D(p+q,M_2^2)}{F(a,z;P,q)}\, .
\ee
 Hence, using the fact that $I_{u}(P,p;a,z)=I_{t}(P,-p;a,z)$ we get
\be
I_{u}(P,p;a,z)=  \int\limits_{-1}^1 d\eta\,
\int\limits_{-\infty}^{\infty}d\alpha\,
\frac{\rho_{t}(\alpha,\eta,a,z)}{(\alpha-M_1^2)F(\al,\eta;P,-p)} \, ,
\ee
which can be written as
\be
I_{u}(P,p;a,z)=  \int\limits_{-1}^1 d\eta\,
\int\limits_{-\infty}^{\infty}d\alpha\,
\frac{\rho_{t}(\alpha,-\eta,a,z)}{(\alpha-M_1^2)F(\al,\eta;P,p)} \, ,
\ee
 since $F(\al,\eta;P,p)=F(\al,-\eta;P,-p)$, or in  other words
$\rho_{u}(\alpha,\eta,a,z)=\rho_{t}(\alpha,-\eta,a,z)$.

%

\end{document}